\documentclass[atmp]{ipart_v1}

\Vol{19}
\Issue{3}
\Year{2015}
\firstpage{701}

\usepackage{t1enc}
\usepackage[latin1]{inputenc}
\usepackage[english]{babel}

\usepackage{amsthm}
\usepackage{yfonts}

\usepackage{bbm}
\usepackage{bm}
\usepackage{mathrsfs}

\newcommand{\be}[0]{\begin{equation}}
\newcommand{\ee}[0]{\end{equation}}

\numberwithin{equation}{section}

\theoremstyle{plain}% default

\begin{document}

\title[Explicit Quantum Green Function for Scattering Problems in 2-D Potential]{Explicit Quantum Green Function for Scattering Problems in 2-D Potential}

\author[Brahim Ben Ali and Mohammed Tayeb Meftah]{Brahim Ben Ali and 
Mohammed Tayeb Meftah}

\begin{abstract}
In this work, we present a new result which concerns the derivation of the Green function relative to the time-independent Schrodinger equation in two dimensional space. The system considered in this work is a quantum particle that have an energy E and moves in an axi- symmetrical potential. Precisely, we have assumed that the potential $V(r)$, in which the quantum particle moves, to be equal to zero inside a disk (radius b) and to be equal a positive constant $V_{0}$ in a crown of internal radius b and external radius $a (b<a)$ and equal zero out side the crown ($r>a$). We have explored the diffusion states regime for which $E>V_{0}$. We have used, to obtain the Green function, the continuity of the solution and of its rst derivative at $r=b$ and $r=a$. We have obtained the associate Green function showing the resonance energies (absence of the reflected waves) for the case $E>V_{0}$.
\end{abstract}

\maketitle

\section{Introduction}

\label{sec1}

The method of the Green function (GF) is a very powerful tool to solve
almost all the problems encountered in mathematical physics, mechanics,
acoustics and electromagnetism.The GF was initially defined in the
distribution theory by Green itself in the electromagnetism theory \cite{gg}%
. Thereafter, (GF) is investigated by other researchers like Neumann \cite%
{neu} in the theory of the Newtonian potential and Helmholtz \cite{helm} in
the theory of acoustics. As for the ordinary differential equations, the
same differential equation can have different GFs according to the initial
conditions and the boundary conditions imposed on the studied problem.
Before starting to expose of our problem, we must specify some references
that deal questions in wide connection with our subject. The authors \cite%
{bc,bc1} have considered the problem of a thin circular plate. They assume
that the plate edge is elastically supported so that the boundary values are
those of the radial bending moment equals zero and the strength is
proportional to the function ofthe deflection on the boundary. In \cite{ku},
the authors examine GF for a circular, annular and exterior circular
domain.In \cite{ki,kiku} the (GF) was obtained for the elliptic domain. \cite%
{ad} treatedhe quantum problem relative to the scatter- ing in two
dimensions. In \cite{nem,nos,lay,tag} the authors, in approximative approach
the (GF) problem was evaluated. In our work, we will interested to
theproblem that consists to compute the GF relative to the Schroedinger
equation in two dimensions: the Shroedinger operator is defined to be
piecewise operator on three connected circular domains ($0<r<b;b<r<a;a<r<%
\infty $) but with specific new boundary conditions. These boundaries
conditions are useful in quantum mechanics to solve the scattering problems
and also the bound states. In quantum mechanics, if the potential is
constant in the crown and is zero outside (or vice versa) the solution of
theSchroedinger equation and the derivative of the solution are continuous
on the boundary (the edge) of the crown. Specify clearly our problem: the
Schrodinger equation takes different forms depending on whether it is inside
the crown ($b<r<a$) or outside. This type of problem matches in quantum
mechanics to the study of a particle subjected to a potential which is a
positive constant inside the crown ($b<r<a$) and zero outside the crown,
that is to say: $r<b$ and $r>a$. None of these cited works, and none to our
knowledge, the explicit Green's function for a piecewise continuous
potential has been calculated in two dimensions for this type of problem.
The physical phenomenon that we want to describe in this work is related to
the resonance phenomenon in one dimension, by extending it to two
dimensions. It is therefore, a question of studying the propagation of the
waves associated with quantum particles (electrons for example) issued from
a source that is located at the space origin, in a homogeneous
two-dimensional medium. During propagation, the particles (waves) enter a
coronal region (barrier) in which they are subjected to a constant potential 
$V_{0}$. Then they cross this region to go to infinity (r tends to
infinity). Another feature of quantum particles, which is not encountered in
classical mechanics, is the well-known the resonance phenomenon in the
scattering regime: when a quantum particle crosses a potential barrier, with
an energy $E>V_{0}$, the probability that the particle reflects is in
general not zero, but it exists certain values of E (resonance energies) for
which there is no reflexion, that is to say there is a total transmission.

So our paper will be organized as it follows: in the next section (Sect.2),
we give a brieve overview on Green's function and its construction, whereas in the third section we expose the problem we will solve. In section three (Sect.4), we will calculate the (GF) for the diffusion regime. It turns out that the resonance
energies are obtained from the poles of the (GF) in the region $r<b $. We
end our paper by a conclusion in Sect.5.

\section{A brieve Green's function overview}

Suppose we have a differential equation of order $n$:
\begin{equation}
L\left[ y\right] \equiv p_{0}\left( x\right) y^{\left( n\right)
}+p_{1}\left( x\right) y^{\left( n-1\right) }+...+p_{n}\left( x\right) y=0
\label{1a}
\end{equation}

where the functions $p_{0}\left( x\right) ,p_{1}\left( x\right)
,...,p_{n}\left( x\right) $ are continuous on $[a,b]$, $p_{0}(x)\neq 0$ on $%
[a,b]$, and the boundary conditions are%
\begin{eqnarray}
V_{k}\left( y\right) &=&\alpha _{k}y\left( a\right) +\alpha
_{k}^{1}y^{\prime }\left( a\right) +...+\alpha _{k}^{n-1}y^{\left(
n-1\right) }+\beta _{k}y\left( b\right) +\beta _{k}^{1}y^{\prime }\left(
b\right) +  \nonumber \\
&&...+\beta _{k}^{n-1}y^{\left( n-1\right) }\left( b\right) ,\text{ \ \ \ \
\ \ }\left( k=1,2,...,n\right) \text{\ }  \label{2a}
\end{eqnarray}

where the linear forms $V_{1},...,V_{n}$ in $y\left( a\right) ,y^{\prime
}\left( a\right) ,...,y^{\left( n-1\right) }\left( a\right) ,y\left(
b\right) ,y^{\prime }\left( b\right) ,...,$

$y^{\left( n-1\right) }\left( b\right) $ are linearly independent.

We assume that the homogeneous boundary-value problem (\ref{1a})-(\ref{2a})
has only a trivial solution $y(x)\equiv 0$.

\textbf{Definition}: Green's function of the boundary-value problem(\ref{1a}%
)-(\ref{2a}) is the function $G(x,\xi )$ constructed for any point $\xi $ such that $a<\xi <b$, and having the following four properties:

(I) $G(x,\xi )$ is continuous and has continuous derivatives with respect to 
$x$ up to order $(n-2)$ inclusive for $a\leq x\leq b$.

(2) Its $(n-1)$th derivative with respect to $x$ at the point $x=\xi $ has a
discontinuity of the first kind, the jump being equal to $\frac{1}{%
p_{0}\left( x\right) }$ , i.e.,%
\begin{equation}
\frac{\partial ^{n-1}G\left( x,\xi _{+}\right) }{\partial x^{n-1}}-\frac{%
\partial ^{n-1}G\left( x,\xi _{-}\right) }{\partial x^{n-1}}=\frac{1}{%
p_{0}\left( x\right) }  \label{3a}
\end{equation}

(3) In each of the intervals $[a,\xi )$ and $(\xi ,b$) the function $G(x,\xi
)$, considered as a function of $x$, is a solution of equation (\ref{1a}):%
\begin{equation}
L\left[ G\right] =0  \label{4a}
\end{equation}

(4) $G(x,\xi )$ satisfies the boundary conditions (\ref{2a}):

\begin{equation}
V_{k}(G)=0,\text{ \ \ \ \ }(k=1,2,...,n)  \label{5a}
\end{equation}
On the exixtence and the unicity of the solution, we refer to the following theorem,\\
\textbf{Theorem} \cite{krasn}: If the boundary-value problem (\ref{1a})-( \ref{2a}) has
only the trivial solution $y(x)=0$, then the operator $L$ has one and only
one Green's function $G(x,\xi )$ (end theorem).\\
It is easy to convince ourselves that the four above conditions are fulfilled and the demonstration of the theorem is in \cite{krasn}. Now we will apply this theorem
to find the Green function for an interesting 2-D problem in quantum physics.

\section{Axi-symmetric two dimensional quantum problem}

Consider a quantum particle moving in a symmetrical potential (independent
of the angle $\theta $) defined as (see fig.1): 
\begin{equation}
V\left( r,\theta \right) =\left\{ 
\begin{array}{c}
0\quad \quad 0\leq r\leq b \\ 
V_{0}\quad \quad b\leq r\leq a \\ 
0\quad \quad \quad \quad r\geq a%
\end{array}%
\right.  \label{1}
\end{equation}

\begin{figure}
%\begin{center}
\includegraphics{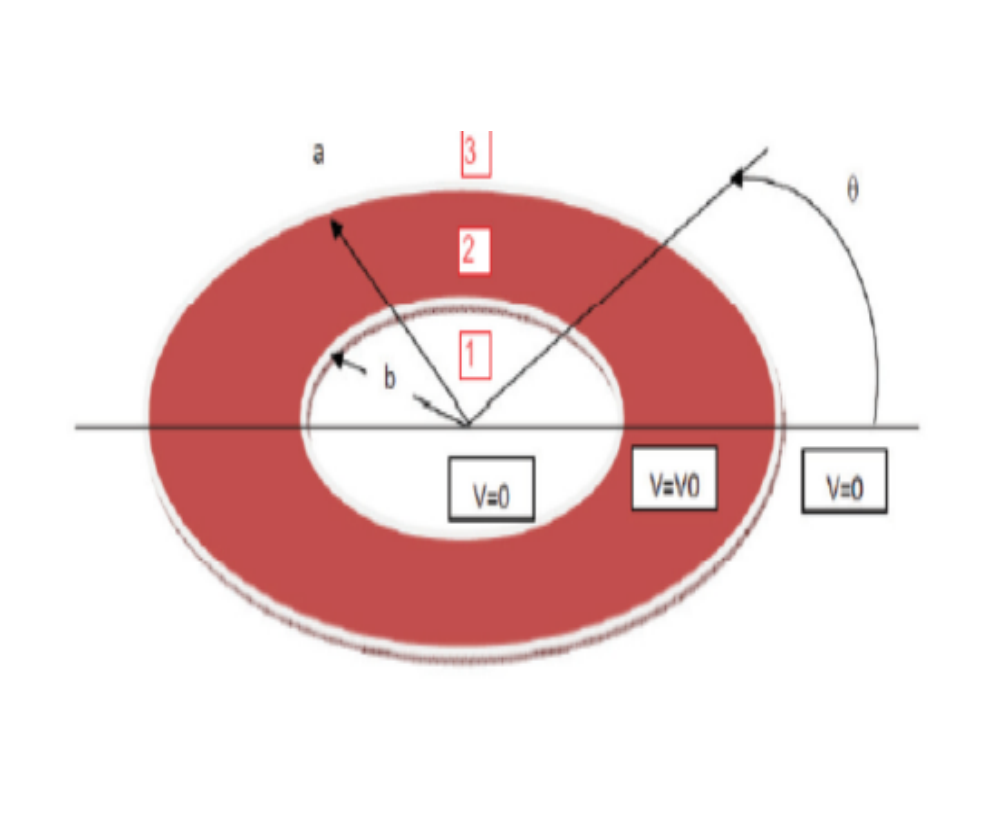}
\caption{A scheme of the coronal potential in two dimensions}
%\end{center}
%\label{fig4}
\end{figure}

The dynamics of this particle is governed by the time-independent
Schroedinger equation:%
\begin{equation}
\hat{H}\left( r,\theta \right) \Psi (r,\theta )=E\Psi \left( r,\theta \right)
\label{a.1}
\end{equation}%
which is written in the natural polar coordinates $\left( r,\theta \right) $
and where $\hat{H}$ $\left( r,\theta \right) $ is the hamiltonien of the
particle, with a mass $M$, moving in this potential. The equation (\ref{a.1}%
) is merely an eigenvalues $E$ and eigenfuntions equation $\Psi \left(
r,\theta \right) $. The explicit form of the hamiltonien of the system is:%
\begin{equation}
\hat{H}=-\frac{\hbar ^{2}}{2M}\triangle _{r,\theta }+V\left( r,\theta \right)
\label{a.2}
\end{equation}%
where: 
\begin{equation}
\triangle _{r,\theta }=\frac{\partial ^{2}}{\partial r^{2}}+\frac{1}{r}\frac{%
\partial }{\partial r}+\frac{1}{r^{2}}\frac{\partial ^{2}}{\partial \theta
^{2}}  \label{13}
\end{equation}%
is the well known laplacian in polar coordinates. The equation ( \ref{a.1} )
writes as:%
\begin{equation}
\left( -\frac{\hbar ^{2}}{2M}\triangle _{r,\theta }+V\left( r,\theta \right)
-E\right) \Psi \left( r,\theta \right) =0  \label{a.4}
\end{equation}%
or, with respect of the definition of $V\left( r,\theta \right) $ in the
formula (\ref{1}) 
\begin{equation}
\left\{ 
\begin{array}{c}
\left( \frac{\hbar ^{2}}{2M}\triangle _{r,\theta }+E\right) \Psi
_{out}\left( r,\theta \right) =0\quad \quad r>a \\ 
\left( \frac{\hbar ^{2}}{2M}\triangle _{r,\theta }-V_{0}+E\right) \Psi
_{mid}\left( r,\theta \right) =0\quad \quad b\leq r\leq a \\ 
\left( \frac{\hbar ^{2}}{2M}\triangle _{r,\theta }+E\right) \Psi
_{int}\left( r,\theta \right) =0\quad \quad \quad \quad 0\leq r\leq b%
\end{array}%
\right.  \label{2}
\end{equation}%
This system is subjected to the boundary conditions defined as $\Psi \left(
r,\theta \right) $ and $\frac{d}{dr}\Psi \left( r,\theta \right) $ are to be
continous at $r=b$ and\ $\ r=a$ for all values of the azimutal angle $\theta 
$. The separation variables method leads to transform the last equations (%
\ref{2}) as 
\begin{equation}
\left\{ 
\begin{array}{c}
\frac{d}{dr}\left( r\frac{d}{dr}\Psi _{out}\right) +\left( \frac{2M}{\hbar
^{2}}Er-\frac{l^{2}}{r}\right) \Psi _{out}(r)=0\quad \quad r>a \\ 
\frac{d}{dr}\left( r\frac{d}{dr}\Psi _{mid}\right) +\left( \frac{2M}{\hbar
^{2}}(E-V_{0})r-\frac{l^{2}}{r}\right) \Psi _{mid}(r)=0\quad \quad b\leq
r\leq a \\ 
\frac{d}{dr}\left( r\frac{d}{dr}\Psi _{int}\right) +\left( \frac{2M}{\hbar
^{2}}Er-\frac{l^{2}}{r}\right) \Psi _{int}(r)=0\quad \quad \quad \quad 0\leq
r\leq b%
\end{array}%
\right.  \label{3}
\end{equation}%
whose solutions are combination of two linear independent Bessel's functions
of order $l$ ($l\in Z$). The solution must obey to the boundary conditions
at $r=b$ and $r=a$:\newline
\begin{align}
\Psi _{out}(a)& =\Psi _{mid}(a) \\
(\frac{d\Psi _{out}(r)}{dr})_{r=a}& =(\frac{d\Psi _{mid}(r)}{dr})_{r=a}
\end{align}%
\newline
and 
\begin{align}
\Psi _{mid}(b)& =\Psi _{int}(b) \\
(\frac{d\Psi _{mid}(r)}{dr})_{r=b}& =(\frac{d\Psi _{int}(r)}{dr})_{r=b}
\end{align}%
\newline
where $l=...-2,-1,0,+1,+2,...$. The global (GF) of the problem (\ref{3})
augmented by the boundary conditions (8-11) is given by 
\begin{equation*}
G(\overrightarrow{r},\overrightarrow{r}^{\prime },E)=G(r,\theta ,r^{\prime
},\theta ^{\prime },E)={\sum\limits_{l=-\infty }^{+\infty }}G(l;r,r^{\prime
},E)\exp (il(\theta -\theta ^{\prime }))
\end{equation*}%
where $G(l;r,r^{\prime },E)\equiv $ $G(l;r,r^{\prime })$ is the radial (GF)
that we shall calculate in the subsequent sections. To calculate the (GF) we
will study separately two cases of the energy: the first case is $E>V_{0}$,
which corresponds to the diffusion regime and the second is $0<E<V_{0}$ for
which corresponds the bounded states regime.

\section{The diffusion states regime: $E>V_{0}$}

\subsection{The region: $a\leq r\leq r^{\prime}<\infty$}

Using the first equation of (\ref{3}), in third region ($r>a$), the
corresponding radial (GF) can be written as the following 
\begin{equation}
G_{out}\left( l;r,r^{\prime }\right) =G^{3,3}\left( l;r,r^{\prime }\right)
=\left\{ 
\begin{array}{c}
C\left( r^{\prime }\right) \left[ Y_{l}\left( kr\right) -\beta (r^{\prime
})J_{l}\left( kr\right) \right] \quad a\leq r\leq r^{\prime } \\ 
D\left( r^{\prime }\right) J_{l}\left( kr\right) \quad \quad r^{\prime }\leq
r<\infty 
\end{array}%
\right.   \label{a..11}
\end{equation}%
where: $k^{2}=\frac{2M}{\hbar ^{2}}E$. Using the continuity of the (GF) at $%
r=r^{\prime }$%
\begin{equation*}
G^{3,3}\left( l;r_{+}^{\prime },r^{\prime }\right) -G^{3,3}\left(
l;r_{-}^{\prime },r^{\prime }\right) =0
\end{equation*}%
then%
\begin{equation}
\left[ D\left( r^{\prime }\right) +\beta (r^{\prime })C\left( r^{\prime
}\right) \right] J_{l}\left( kr^{\prime }\right) -C\left( r^{\prime }\right)
Y_{l}\left( kr^{\prime }\right) =0  \label{a..12}
\end{equation}%
and the dicontinuity of the first derivative with respect $r$ at $%
r=r^{\prime }$:%
\begin{equation*}
\frac{d}{dr}G^{3,3}\left( l;r_{+}^{\prime },r^{\prime }\right) -\frac{d}{dr}%
G^{3,3}\left( l;r_{-}^{\prime },r^{\prime }\right) =\frac{2}{\pi r^{\prime }}
\end{equation*}%
then%
\begin{equation}
\left[ D\left( r^{\prime }\right) +\beta (r^{\prime })C\left( r^{\prime
}\right) \right] J_{l}^{\prime }\left( kr^{\prime }\right) -C\left(
r^{\prime }\right) Y_{l}^{\prime }\left( kr^{\prime }\right) =\frac{2}{\pi
kr^{\prime }}.  \label{a13}
\end{equation}%
By comparinng ( \ref{a..12} ) and (\ref{a13}) we check that 
\begin{equation}
C\left( r^{\prime }\right) \left[ Y_{l}\left( kr^{\prime }\right)
J_{l}^{\prime }\left( kr^{\prime }\right) -J_{l}\left( kr^{\prime }\right)
Y_{l}^{\prime }\left( kr^{\prime }\right) \right] =\frac{2J_{l}\left(
kr^{\prime }\right) }{\pi kr^{\prime }}  \label{a..14}
\end{equation}%
and by using the Bessel Wronskian for the pair $\left( J_{l},Y_{l}\right) $:%
\begin{equation}
W\left( J_{l}\left( kr^{\prime }\right) ,Y_{l}\left( kr^{\prime }\right)
\right) =J_{l}\left( kr^{\prime }\right) Y_{l}^{\prime }\left( kr^{\prime
}\right) -J_{l}^{\prime }\left( kr^{\prime }\right) Y_{l}\left( kr^{\prime
}\right) =\frac{2}{\pi kr^{\prime }}  \label{a..15}
\end{equation}%
it is easy to get: 
\begin{equation}
C\left( r^{\prime }\right) =-J_{l}\left( kr^{\prime }\right)   \label{19aa}
\end{equation}%
and then from (\ref{a13}) we obtain:

\begin{equation}
D\left( r^{\prime }\right) =\beta (r^{\prime })J_{l}\left( kr^{\prime
}\right) -Y_{l}\left( kr^{\prime }\right) .  \label{a..16}
\end{equation}%
After substitution of (\ \ref{a..16} ) and ( \ref{19aa} ) in ( \ref{a..11} )
we find the (GF) in the region ( $r,r^{\prime }\geq a$): 
\begin{equation}
G^{3,3}\left( l;r,r^{\prime }\right) =-\left\{ 
\begin{array}{c}
J_{l}\left( kr^{\prime }\right) \left[ Y_{l}\left( kr\right) -\beta
(r^{\prime })J_{l}\left( kr\right) \right] \quad \quad a\leq r\leq r^{\prime
} \\ 
\left[ Y_{l}\left( kr^{\prime }\right) -\beta (r^{\prime })J_{l}\left(
kr^{\prime }\right) \right] J_{l}\left( kr\right) \quad \quad r^{\prime
}\leq r<\infty 
\end{array}%
\right. .  \label{a..17}
\end{equation}%
It remains to determine the coefficient $\beta (r^{\prime })$. To do this,
we use the symmetry property\newline
\begin{equation*}
G^{3,3}(l:r,r^{\prime })=G^{3,3}(l:r^{\prime },r)
\end{equation*}%
then 
\begin{equation*}
\left[ Y_{l}\left( kr^{\prime }\right) -\beta (r^{\prime })J_{l}\left(
kr^{\prime }\right) \right] J_{l}\left( kr\right) =\left[ Y_{l}\left(
kr\right) -\beta (r)J_{l}\left( kr\right) \right] J_{l}\left( kr^{\prime
}\right) 
\end{equation*}%
By identifying in the last equation we find

\begin{equation}
\beta (r^{\prime })=\beta (r)=\beta 
\end{equation}%
Then the (GF) in this region ( $r,r^{\prime }\geq a$) is given by:%
\begin{equation}
G^{3,3}\left( l;r,r^{\prime }\right) =-\left\{ 
\begin{array}{c}
J_{l}\left( kr^{\prime }\right) \left[ Y_{l}\left( kr\right) -\beta
J_{l}\left( kr\right) \right] \quad \quad a\leq r\leq r^{\prime } \\ 
\left[ Y_{l}\left( kr^{\prime }\right) -\beta J_{l}\left( kr^{\prime
}\right) \right] J_{l}\left( kr\right) \quad \quad r^{\prime }\leq r<\infty 
\end{array}%
\right.   \label{112}
\end{equation}%
where $\beta $ is a constant to be determined lates.

\subsection{The region: $b\leq r\leq r^{\prime }\leq a$\protect\bigskip\ }

The Green's function in this region can be written as:%
\begin{equation*}
G_{mid}(l;r,r^{\prime })=G^{2,2}\left( l;r,r^{\prime }\right) =\left\{ 
\begin{array}{c}
E\left( r^{\prime }\right) \left[ Y_{l}\left( \mu r\right) -\delta
(r^{\prime })J_{l}\left( \mu r\right) \right] \quad \quad b\leq r\leq
r^{\prime } \\ 
F\left( r^{\prime }\right) \left[ Y_{l}\left( \mu r\right) -\gamma \left(
r^{\prime }\right) J_{l}\left( \mu r\right) \right] \quad \quad r^{\prime
}\leq r<a%
\end{array}%
\right. 
\end{equation*}%
where: $\mu ^{2}=\frac{2M}{\hbar ^{2}}(E-V_{0})$. To calculate the
coefficients $E\left( r^{\prime }\right) $, $F\left( r^{\prime }\right) $, $%
\gamma (r^{\prime })$ and $\delta (r^{\prime })$ we use the continuity of
the (GF) at $r=r^{\prime }$:%
\begin{equation*}
G^{2,2}\left( l;r_{+}^{\prime },r^{\prime }\right) =G^{2,2}\left(
l;r_{-}^{\prime },r^{\prime }\right) 
\end{equation*}%
then%
\begin{equation}
Y_{l}\left( \mu r^{\prime }\right) \left[ F\left( r^{\prime }\right)
-E\left( r^{\prime }\right) \right] -J_{l}\left( \mu r^{\prime }\right) %
\left[ \gamma \left( r^{\prime }\right) F\left( r^{\prime }\right) -\delta
(r^{\prime })E\left( r^{\prime }\right) \right] =0  \label{aa2}
\end{equation}%
and we use the discontinuity of the first derivative with respect $r$ \ at $%
r=r^{\prime }$:%
\begin{equation*}
\frac{d}{dr}G^{2,2}\left( l;r=r_{+}^{\prime },r^{\prime }\right) -\frac{d}{dr%
}G^{2,2}\left( l;r=r_{-}^{\prime },r^{\prime }\right) =\frac{2}{\pi
r^{\prime }}
\end{equation*}%
then%
\begin{equation}
Y_{l}^{\prime }\left( \mu r^{\prime }\right) \left[ F\left( r^{\prime
}\right) -E\left( r^{\prime }\right) \right] -J_{l}^{\prime }\left( \mu
r^{\prime }\right) \left[ \gamma \left( r^{\prime }\right) F\left( r^{\prime
}\right) -\delta (r^{\prime })E\left( r^{\prime }\right) \right] =\frac{2}{%
\pi \mu r^{\prime }}.  \label{aa3}
\end{equation}%
By combining ( \ref{aa2} ) and ( \ref{aa3} ), we obtain:%
\begin{equation*}
F\left( r^{\prime }\right) =\frac{E\left( r^{\prime }\right) \left[
Y_{l}\left( \mu r^{\prime }\right) -\delta \left( r^{\prime }\right)
J_{l}\left( \mu r^{\prime }\right) \right] }{\left[ Y_{l}\left( \mu
r^{\prime }\right) -\gamma \left( r^{\prime }\right) J_{l}\left( \mu
r^{\prime }\right) \right] }
\end{equation*}%
and%
\begin{equation*}
Y_{l}^{\prime }\left( \mu r^{\prime }\right) \left[ \frac{E\left( r^{\prime
}\right) \left[ Y_{l}\left( \mu r^{\prime }\right) -\delta \left( r^{\prime
}\right) J_{l}\left( \mu r^{\prime }\right) \right] }{\left[ Y_{l}\left( \mu
r^{\prime }\right) -\gamma \left( r^{\prime }\right) J_{l}\left( \mu
r^{\prime }\right) \right] }-E\left( r^{\prime }\right) \right] 
\end{equation*}%
\begin{equation*}
-J_{l}^{\prime }\left( \mu r^{\prime }\right) \left[ \gamma \left( r^{\prime
}\right) \frac{E\left( r^{\prime }\right) \left[ Y_{l}\left( \mu r^{\prime
}\right) -\delta \left( r^{\prime }\right) J_{l}\left( \mu r^{\prime
}\right) \right] }{\left[ Y_{l}\left( \mu r^{\prime }\right) -\gamma \left(
r^{\prime }\right) J_{l}\left( \mu r^{\prime }\right) \right] }-\delta
(r^{\prime })E\left( r^{\prime }\right) \right] =\frac{2}{\pi \mu r^{\prime }%
}
\end{equation*}%
Using the Bessel Wronksian for the pair $\left( J_{l},Y_{l}\right) $:%
\begin{equation}
W\left( J_{l}\left( \mu r^{\prime }\right) ,Y_{l}\left( \mu r^{\prime
}\right) \right) =J_{l}\left( \mu r^{\prime }\right) Y_{l}^{\prime }\left(
\mu r^{\prime }\right) -Y_{l}\left( \mu r^{\prime }\right) J_{l}^{\prime
}\left( \mu r^{\prime }\right) =\frac{2}{\pi \mu r^{\prime }}
\end{equation}%
we get the coefficients:%
\begin{equation}
E\left( r^{\prime }\right) =\frac{Y_{l}\left( \mu r^{\prime }\right) -\gamma
\left( r^{\prime }\right) J_{l}\left( \mu r^{\prime }\right) }{g(r^{\prime })%
}
\end{equation}%
where 
\begin{equation}
g(x)=\gamma \left( x\right) -\delta \left( x\right) 
\end{equation}%
and%
\begin{equation}
F\left( r^{\prime }\right) =\frac{Y_{l}\left( \mu r^{\prime }\right) -\delta
\left( r^{\prime }\right) J_{l}\left( \mu r^{\prime }\right) }{g(r^{\prime })%
}
\end{equation}%
Then, the (GF) in the region $b\leq r\leq r^{\prime }\leq a$ is given by:\\
$
G^{2,2}\left( l;r,r^{\prime }\right) =\frac{1}{g(r^{\prime })}\left\{ 
\begin{array}{c}
\left[ Y_{l}\left( \mu r^{\prime }\right) -\gamma \left( r^{\prime }\right)
J_{l}\left( \mu r^{\prime }\right) \right] \left[ Y_{l}\left( \mu r\right)
-\delta (r^{\prime })J_{l}\left( \mu r\right) \right]  \\ 
\left[ Y_{l}\left( \mu r^{\prime }\right) -\delta \left( r^{\prime }\right)
J_{l}\left( \mu r^{\prime }\right) \right] \left[ Y_{l}\left( \mu r\right)
-\gamma \left( r^{\prime }\right) J_{l}\left( \mu r\right) \right] 
\end{array}%
\right. $

for $b\leq r\leq r^{\prime }\leq a$ and $b\leq r^{\prime }\leq r\leq a$
respectively. It remains to determine the coefficients $\delta (r^{\prime }),
$ $\gamma (r^{\prime })$ and $g(r^{\prime })$. To do this, we use the
symmetry property of $G(l;r,r^{\prime })$%
\begin{equation*}
G^{2,2}(l;r,r^{\prime })=G^{2,2}(l;r^{\prime },r)
\end{equation*}%
then%
\begin{equation*}
\frac{1}{g(r^{\prime })}\left\{ 
\begin{array}{c}
\left[ Y_{l}\left( \mu r^{\prime }\right) -\gamma \left( r^{\prime }\right)
J_{l}\left( \mu r^{\prime }\right) \right] \left[ Y_{l}\left( \mu r\right)
-\delta (r^{\prime })J_{l}\left( \mu r\right) \right]  \\ 
\left[ Y_{l}\left( \mu r^{\prime }\right) -\delta \left( r^{\prime }\right)
J_{l}\left( \mu r^{\prime }\right) \right] \left[ Y_{l}\left( \mu r\right)
-\gamma \left( r^{\prime }\right) J_{l}\left( \mu r\right) \right] 
\end{array}%
\right. 
\end{equation*}%
\begin{equation*}
=\frac{1}{g(r)}\left\{ 
\begin{array}{c}
\left[ Y_{l}\left( \mu r\right) -\gamma \left( r\right) J_{l}\left( \mu
r\right) \right] \left[ Y_{l}\left( \mu r^{\prime }\right) -\delta
(r)J_{l}\left( \mu r^{\prime }\right) \right]  \\ 
\left[ Y_{l}\left( \mu r\right) -\delta \left( r\right) J_{l}\left( \mu
r\right) \right] \left[ Y_{l}\left( \mu r^{\prime }\right) -\gamma \left(
r\right) J_{l}\left( \mu r^{\prime }\right) \right] 
\end{array}%
\right. 
\end{equation*}%
By identifying in the last equation we find%
\begin{align}
\delta \left( r\right) & =\delta \left( r^{\prime }\right) =\delta =constant
\\
\gamma \left( r\right) & =\gamma \left( r^{\prime }\right) =\gamma =constant
\notag \\
g\left( r\right) & =g\left( r^{\prime }\right) =g=constant  \label{111}
\end{align}%
These constants we must to determine later.

\subsection{The coefficients $\protect\gamma$ and $\protect\delta$
determination}

To find the coefficients $\gamma$ and $\delta$ we use the continuity of the
(GF) and the continuity of its derivative at $r=a$:

\begin{equation*}
G^{3,3}\left( l;r,a\right) =G^{2,2}\left( l;r,a\right)
\end{equation*}
then%
\begin{equation}
\frac{1}{g}\left[ Y_{l}\left( \mu a\right) -\gamma J_{l}\left( \mu a\right) %
\right] \left[ Y_{l}\left( \mu a\right) -\delta J_{l}\left( \mu a\right) %
\right] =-J_{l}\left( ka\right) \left[ Y_{l}\left( ka\right) -\beta
J_{l}\left( ka\right) \right]  \label{114}
\end{equation}
and%
\begin{equation*}
\frac{d}{dr}G^{3,3}\left( l;r,a\right) \rfloor_{r=a}=\frac{d}{dr}%
G^{2,2}\left( l;r,a\right) \rfloor_{r=a}
\end{equation*}
then

\begin{equation}
\frac{\mu }{g}\left[ Y_{l}^{\prime }\left( \mu a\right) -\gamma
J_{l}^{\prime }\left( \mu a\right) \right] \left[ Y_{l}\left( \mu a\right)
-\delta J_{l}\left( \mu a\right) \right] =kJ_{l}\left( ka\right) \left[
\beta J_{l}^{\prime }\left( ka\right) -Y_{l}^{\prime }\left( ka\right) %
\right]   \label{113}
\end{equation}%
By dividing (\ref{113}) by (\ref{114}) and after simplifications we get the
following equation

\begin{equation*}
\frac{\mu \left[ Y_{l}^{\prime }\left( \mu a\right) -\gamma J_{l}^{\prime
}\left( \mu a\right) \right] }{\left[ Y_{l}\left( \mu a\right) -\gamma
J_{l}\left( \mu a\right) \right] }=\frac{k\left[ Y_{l}^{\prime }\left(
ka\right) -\beta J_{l}^{\prime }\left( ka\right) \right] }{\left[
Y_{l}\left( ka\right) -\beta J_{l}\left( ka\right) \right] }
\end{equation*}%
from which we get the coefficient $\gamma $ 
\begin{align}
\gamma & =\frac{kY_{l}\left( \mu a\right) \left[ Y_{l}^{\prime }\left(
ka\right) -\beta J_{l}^{\prime }\left( ka\right) \right] -\mu Y_{l}^{\prime
}\left( \mu a\right) \left[ Y_{l}\left( ka\right) -\beta J_{l}\left(
ka\right) \right] }{kJ_{l}\left( \mu a\right) \left[ Y_{l}^{\prime }\left(
ka\right) -\beta J_{l}^{\prime }\left( ka\right) \right] -\mu J_{l}^{\prime
}\left( \mu a\right) \left[ Y_{l}\left( ka\right) -\beta J_{l}\left(
ka\right) \right] } \\
& or  \notag \\
& \gamma (k,\mu ,a,b)=\frac{V(k,\mu ,a,b)}{U(k,\mu ,a,b)}  \label{116}
\end{align}%
such that%
\begin{align}
V(k,\mu ,\beta ,a,b)& =kY_{l}\left( \mu a\right) \left[ Y_{l}^{\prime
}\left( ka\right) -\beta J_{l}^{\prime }\left( ka\right) \right]   \notag \\
& -\mu Y_{l}^{\prime }\left( \mu a\right) \left[ Y_{l}\left( ka\right)
-\beta J_{l}\left( ka\right) \right]   \label{a1a}
\end{align}%
\begin{align}
U(k,\mu ,\beta ,a,b)& =kJ_{l}\left( \mu a\right) \left[ Y_{l}^{\prime
}\left( ka\right) -\beta J_{l}^{\prime }\left( ka\right) \right]   \notag \\
& -\mu J_{l}^{\prime }\left( \mu a\right) \left[ Y_{l}\left( ka\right)
-\beta J_{l}\left( ka\right) \right] .  \label{a2a}
\end{align}%
By combining ( \ref{114} ) and ( \ref{116} ) we obtain:

\begin{equation*}
\delta =\frac{\mu Y_{l}\left( \mu a\right) \left[ Y_{l}^{\prime }\left( \mu
a\right) -\gamma J_{l}^{\prime }\left( \mu a\right) \right] +k\gamma
J_{l}\left( ka\right) \left[ Y_{l}^{\prime }\left( ka\right) -\beta
J_{l}^{\prime }\left( ka\right) \right] }{\mu J_{l}\left( \mu a\right) \left[
Y_{l}^{\prime }\left( \mu a\right) -\gamma J_{l}^{\prime }\left( \mu
a\right) \right] +kJ_{l}\left( ka\right) \left[ Y_{l}^{\prime }\left(
ka\right) -\beta J_{l}^{\prime }\left( ka\right) \right] }
\end{equation*}%
\begin{equation*}
\delta \equiv \delta (k,\mu ,a,b)=\frac{2Y_{l}\left( \mu a\right) +\pi
aJ_{l}\left( ka\right) V(k,\mu ,a)}{2J_{l}\left( \mu a\right) +\pi
aJ_{l}\left( ka\right) U(k,\mu ,a)}.
\end{equation*}%
Then%
\begin{equation*}
g\equiv g(k,\mu ,a,b)=\gamma (k,\mu ,a,b)-\delta (k,\mu ,a,b)
\end{equation*}%
but $g$ still depends on $\beta $ via the above expressions of $\gamma $ and 
$\delta $ themselves via V and U. We will show later that this dependence
will be removed by showing that $\beta $ also depends on $k,\mu ,a$ and $b.$
With the same way, we find finally, the (GF) (31) in the region $\left(
b\leq r\leq a\right) $ 
\begin{equation*}
G^{2,2}\left( l;r,r^{\prime }\right) =\left\{ 
\begin{array}{c}
\left[ \frac{Y_{l}\left( \mu r^{\prime }\right) }{g(k,\mu ,a,b)}-\frac{%
V(k,\mu ,a,b)}{U(k,\mu ,a,b)}\frac{J_{l}\left( \mu r^{\prime }\right) }{%
g(k,\mu ,a)}\right] \times  \\ 
\left[ Y_{l}\left( \mu r\right) -\frac{2Y_{l}\left( \mu a\right) +\pi
aJ_{l}\left( ka\right) V(k,\mu ,a,b)}{2J_{l}\left( \mu a\right) +\pi
aJ_{l}\left( ka\right) U(k,\mu ,a,b)}J_{l}\left( \mu r\right) \right]  \\ 
r\longleftrightarrow r^{\prime }%
\end{array}%
\right. 
\end{equation*}%
for $b\leq r\leq r^{\prime }\leq a$ and $a\leq r^{\prime }\leq r\leq a$
respectively.

\subsection{The region:\textbf{\ }$0\leq r\leq r^{\prime}\leq b$}

In this region, the (GF) can be written as:

\begin{equation}
G_{in}\left( l;r,r^{\prime }\right) \equiv G^{1,1}\left( l;r,r^{\prime
}\right) =\left\{ 
\begin{array}{c}
A\left( r^{\prime }\right) J_{l}\left( kr\right) \quad \quad \quad 0<r\leq
r^{\prime } \\ 
B\left( r^{\prime }\right) \left[ Y_{l}\left( kr\right) -\alpha (r^{\prime
})J_{l}\left( kr\right) \right] \quad r^{\prime }\leq r\leq b%
\end{array}%
\right.   \label{a..1}
\end{equation}%
where $k^{2}=\frac{2M}{\hbar ^{2}}E$. To calculate the coefficients $A\left(
r^{\prime }\right) $, $B\left( r^{\prime }\right) $ and $\alpha (r^{\prime })
$, we use the continuity of the (GF) at $r=r^{\prime }$:%
\begin{equation*}
G^{1,1}\left( l;r_{+}^{\prime },r^{\prime }\right) -G^{1,1}\left(
l;r_{-}^{\prime },r^{\prime }\right) =0
\end{equation*}%
then%
\begin{equation}
B\left( r^{\prime }\right) Y_{l}\left( kr^{\prime }\right) -\left[ A\left(
r^{\prime }\right) +\alpha (r^{\prime })B\left( r^{\prime }\right) \right]
J_{l}\left( kr^{\prime }\right) =0  \label{.a.2}
\end{equation}%
and we use the discontinuity of the first derivative with respect $r$ at $%
r=r^{\prime }$ 
\begin{equation*}
\frac{d}{dr}G^{1,1}\left( l;r_{+}^{\prime },r^{\prime }\right) -\frac{d}{dr}%
G^{1,1}\left( l;r_{-}^{\prime },r^{\prime }\right) =\frac{2}{\pi r^{\prime }}
\end{equation*}%
then%
\begin{equation}
B\left( r^{\prime }\right) Y_{l}^{\prime }\left( kr^{\prime }\right) -\left[
A\left( r^{\prime }\right) +\alpha (r^{\prime })B\left( r^{\prime }\right) %
\right] J_{l}^{\prime }\left( kr^{\prime }\right) =\frac{2}{\pi kr^{\prime }}
\label{aa..3}
\end{equation}%
By combining ( \ref{a111} ) and ( \ref{az} ) we obtain 
\begin{equation}
A\left( r^{\prime }\right) =\frac{B\left( r^{\prime }\right) \left[
Y_{l}\left( kr^{\prime }\right) -\alpha (r^{\prime })J_{l}\left( kr^{\prime
}\right) \right] }{J_{l}\left( kr^{\prime }\right) }  \label{a..4}
\end{equation}%
and%
\begin{equation}
B\left( r^{\prime }\right) Y_{l}^{\prime }\left( kr^{\prime }\right) -\left[ 
\frac{B\left( r^{\prime }\right) \left[ Y_{l}\left( kr^{\prime }\right)
-\alpha (r^{\prime })J_{l}\left( kr^{\prime }\right) \right] }{J_{l}\left(
kr^{\prime }\right) }+\alpha (r^{\prime })B\left( r^{\prime }\right) \right]
J_{l}^{\prime }\left( kr^{\prime }\right) =\frac{2}{\pi kr^{\prime }}
\label{a..5}
\end{equation}%
By using the Bessel Wronksian for the pair $\left( J_{l},Y_{l}\right) $ 
\begin{equation}
W\left( J_{l}\left( kr^{\prime }\right) ,Y_{l}\left( kr^{\prime }\right)
\right) =J_{l}\left( kr^{\prime }\right) Y_{l}^{\prime }\left( kr^{\prime
}\right) -Y_{l}\left( kr^{\prime }\right) J_{l}^{\prime }\left( kr^{\prime
}\right) =\frac{2}{\pi kr^{\prime }}  \label{a..7}
\end{equation}%
we get the coefficients 
\begin{equation}
B\left( r^{\prime }\right) =J_{l}\left( kr^{\prime }\right)   \label{a..8}
\end{equation}%
and%
\begin{equation}
A\left( r^{\prime }\right) =\left[ Y_{l}\left( kr^{\prime }\right) -\alpha
(r^{\prime })J_{l}\left( kr^{\prime }\right) \right]   \label{a..9}
\end{equation}%
Then, the (GF) in this region $\left( r\leq b\right) $ is given by:%
\begin{equation}
G^{1,1}\left( l;r,r^{\prime }\right) =\left\{ 
\begin{array}{c}
\left[ Y_{l}\left( kr^{\prime }\right) -\alpha (r^{\prime })J_{l}\left(
kr^{\prime }\right) \right] J_{l}\left( kr\right) \qquad 0<r\leq r^{\prime }
\\ 
\left[ Y_{l}\left( kr\right) -\alpha (r^{\prime })J_{l}\left( kr\right) %
\right] J_{l}\left( kr^{\prime }\right) \qquad r^{\prime }\leq r\leq b%
\end{array}%
\right.   \label{a..10}
\end{equation}%
It remains to determine the coefficient $\alpha (r^{\prime })$. To do this,
we use the symmetry property of $G(l:r,r^{\prime })$%
\begin{equation*}
G^{1,1}(l;r,r^{\prime })=G^{1,1}(l;r^{\prime },r)
\end{equation*}%
\begin{equation*}
\left[ Y_{l}\left( kr^{\prime }\right) -\alpha (r^{\prime })J_{l}\left(
kr^{\prime }\right) \right] J_{l}\left( kr\right) =\left[ Y_{l}\left(
kr\right) -\alpha (r)J_{l}\left( kr\right) \right] J_{l}\left( kr^{\prime
}\right) 
\end{equation*}%
By identifying in the last equation we find

\begin{equation}
\alpha(r^{\prime})=\alpha(r)=\alpha
\end{equation}
Then the (GF) in this region$\left( r\leq b\right) $ is given by:

\begin{equation}
G^{1,1}\left( l;r,r^{\prime}\right) =\left\{ 
\begin{array}{c}
\left[ Y_{l}\left( kr^{\prime}\right) -\alpha J_{l}\left( kr^{\prime
}\right) \right] J_{l}\left( kr\right) \qquad0<r\leq r^{\prime} \\ 
\left[ Y_{l}\left( kr\right) -\alpha J_{l}\left( kr\right) \right]
J_{l}\left( kr^{\prime}\right) \qquad r^{\prime}\leq r\leq b%
\end{array}
\right.
\end{equation}
We mention here that the coefficient $\alpha$ will be determined in the next
subsection.

\subsection{\protect\bigskip The coefficient $\protect\alpha$ determination}

To find the coefficient $\alpha$, we use the continuity of the (GF) and the
continuity of its derivative at $r=b$ :\ 
\begin{equation*}
G^{1,1}\left( l;r,b\right) =G^{2,2}\left( l;r,b\right)
\end{equation*}
then%
\begin{align}
\left[ Y_{l}\left( \mu b\right) -\frac{2Y_{l}\left( \mu a\right) +\pi
aJ_{l}\left( ka\right) V(k,\mu,a,b)}{2J_{l}\left( \mu a\right) +\pi
aJ_{l}\left( ka\right) U(k,\mu,a,b)}J_{l}\left( \mu b\right) \right] \times &
\notag \\
\left[ \frac{Y_{l}\left( \mu b\right) }{g(k,\mu,a,b)}-\frac{V(k,\mu ,a,b)}{%
U(k,\mu,a,b)}\frac{J_{l}\left( \mu b\right) }{g(k,\mu,a,b)}\right] =[\alpha
J_{l}( kb) -Y_{l}( kb)] J_{l}\left( kb\right)  \label{az}
\end{align}
and using 
\begin{equation*}
\frac{d}{dr}G^{1,1}\left( l;r,b\right) \rfloor_{r=b}=\frac{d}{dr}%
G^{2,2}\left( l;r,b\right) \rfloor_{r=b}
\end{equation*}
or equivalently 
\begin{equation*}
\left[ Y_{l}^{\prime}\left( \mu b\right) -\frac{2Y_{l}\left( \mu a\right)
+\pi aJ_{l}\left( ka\right) V(k,\mu,a,b)}{2J_{l}\left( \mu a\right) +\pi
aJ_{l}\left( ka\right) U(k,\mu,a,b)}J_{l}^{\prime}\left( \mu b\right) \right]
\times
\end{equation*}
\begin{align}
\frac{\mu}{g(k,\mu,a,b)}\left[ Y_{l}\left( \mu b\right) -\gamma J_{l}\left(
\mu b\right) \right] =-k\left[ Y_{l}\left( kb\right) -\alpha J_{l}\left(
kb\right) \right] J_{l}^{\prime}\left( kb\right)  \label{3a34}
\end{align}
By deviding (\ref{3a34}) by (\ref{az}) we find 
\begin{equation}
\frac{\mu\left[ Y_{l}^{\prime}\left( \mu b\right) -\frac{2Y_{l}\left( \mu
a\right) +\pi aJ_{l}\left( ka\right) V(k,\mu,a,b)}{2J_{l}\left( \mu a\right)
+\pi aJ_{l}\left( ka\right) U(k,\mu,a,b)}J_{l}^{\prime}\left( \mu b\right) %
\right] }{\left[ Y_{l}\left( \mu b\right) -\frac{2Y_{l}\left( \mu a\right)
+\pi aJ_{l}\left( ka\right) V(k,\mu,a,b)}{2J_{l}\left( \mu a\right) +\pi
aJ_{l}\left( ka\right) U(k,\mu,a,b)}J_{l}\left( \mu b\right) \right] }=\frac{%
kJ_{l}^{\prime}\left( kb\right) }{J_{l}\left( kb\right) }  \label{3.a}
\end{equation}
After replacing $U$ and $V$ in (\ref{3.a}) by their expressions (\ref{a1a})
et (\ref{a2a}) we find 
\begin{equation}
\beta(k,\mu,a,b)=\frac{F(k,\mu,a,b)}{T(k,\mu,a,b)}  \label{a33}
\end{equation}
such that 
\begin{align}
F(k,\mu,a,b) =\mu J_{l}\left( kb\right) \left[ 2+k\pi aJ_{l}\left( ka\right)
Y_{l}^{\prime}\left( ka\right) \right] F_{1}(k,\mu ,a,b)+  \notag \\
kJ_{l}^{\prime}\left( kb\right) \left[ 2+k\pi aJ_{l}\left( ka\right)
J_{l}^{\prime}\left( ka\right) \right] T_{1}(k,\mu,a,b)+  \notag
\end{align}%
\begin{align}
\mu\pi aJ_{l}\left( ka\right) Y_{l}\left( ka\right) \left[ \mu J_{l}\left(
kb\right) T_{2}(k,\mu,a,b)+kJ_{l}^{\prime}\left( kb\right) F_{2}(k,\mu,a,b)%
\right]  \label{a333}
\end{align}
and 
\begin{align}
T(k,\mu,a,b) =\pi ak\mu J_{l}\left( ka\right) \times  \notag \\
\left[ J_{l}^{\prime}\left( ka\right) J_{l}\left( kb\right) F_{1}
(k,\mu,a,b)-J_{l}\left( ka\right) J_{l}^{\prime}\left( kb\right)
F_{2}(k,\mu,a,b)\right]  \notag
\end{align}
\begin{align}
-\pi aJ_{l}\left( ka\right) \left[ \mu^{2}J_{l}\left( ka\right) J_{l}\left(
kb\right) T_{2}(k,\mu,a,b)+k^{2}J_{l}^{\prime}\left( ka\right)
J_{l}^{\prime}\left( kb\right) T_{1}(k,\mu,a,b)\right]  \label{a334}
\end{align}
where%
\begin{align}
F_{1}(k,\mu,a,b)=J_{l}\left( \mu a\right) Y_{l}^{\prime}\left( \mu b\right)
-Y_{l}\left( \mu a\right) J_{l}^{\prime}\left( \mu b\right) \\
F_{2}(k,\mu,a,b)=J_{l}\left( \mu b\right) Y_{l}^{\prime}\left( \mu a\right)
-Y_{l}\left( \mu b\right) J_{l}^{\prime}\left( \mu a\right) \\
T_{1}(k,\mu,a,b)=J_{l}\left( \mu a\right) Y_{l}\left( \mu b\right)
-Y_{l}\left( \mu a\right) J_{l}\left( \mu b\right) \\
T_{2}(k,\mu,a,b)=J_{l}^{\prime}\left( \mu b\right) Y_{l}^{\prime}\left( \mu
a\right) -Y_{l}^{\prime}\left( \mu b\right) J_{l}^{\prime}\left( \mu a\right)
\end{align}
By using \ref{.a.2} and \ref{.a.2} and after a minor simplifications we get
the coefficient $\alpha$ equal to

\begin{figure}
\begin{center}
\includegraphics{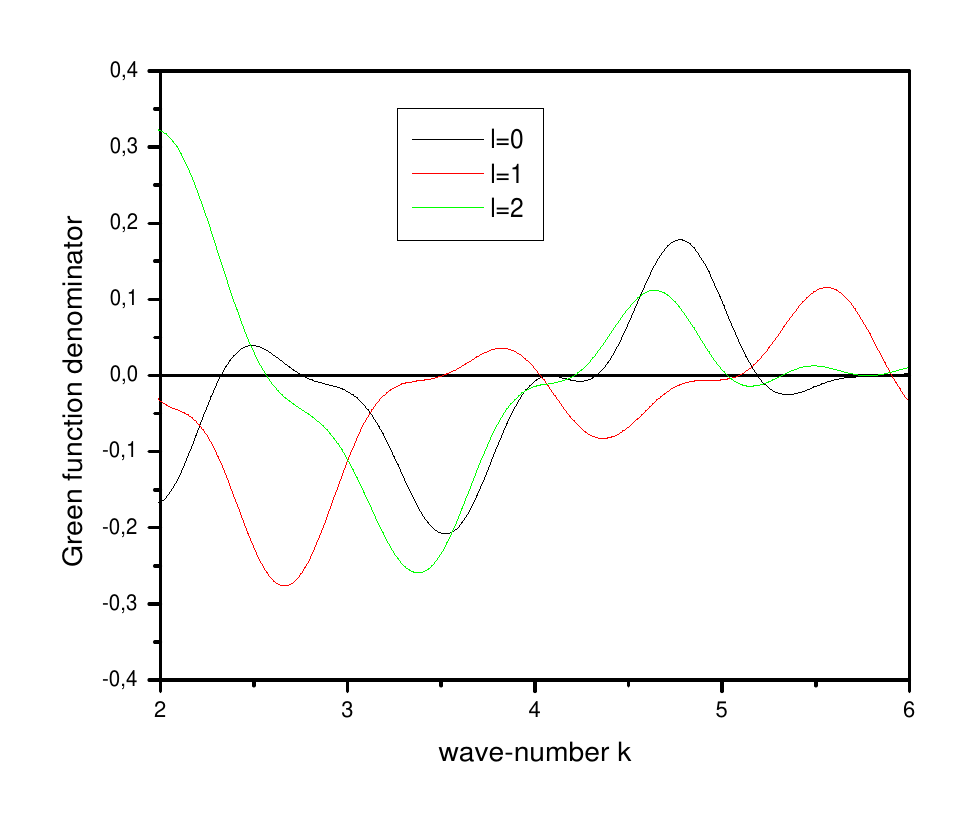}
\caption{The resonance energies are given by the intersection of the curve with k axis for various angular momenta l=0,1,2} 
\end{center}
\label{fig2}
\end{figure}

\begin{equation*}
\alpha (k,\mu ,a,b)=\frac{Y_{l}\left( kb\right) }{J_{l}\left( kb\right) }+%
\left[ \frac{Y_{l}\left( \mu b\right) }{g(k,\mu ,a,b)}-\frac{V(k,\mu ,\beta
,a,b)}{U(k,\mu ,\beta ,a,b)}\frac{J_{l}\left( \mu b\right) }{g(k,\mu ,a,b)}%
\right] \times \newline
\end{equation*}%
\begin{equation*}
\frac{\left[ Y_{l}\left( \mu b\right) -\delta (k,\mu ,a,b)J_{l}\left( \mu
b\right) \right] }{J_{l}^{2}\left( kb\right) }=\frac{Y_{l}\left( kb\right) }{%
J_{l}\left( kb\right) }+\psi (k,\mu ,\ beta,a,b)
\end{equation*}%
such that 
\begin{equation*}
\psi (k,\mu ,\beta ,a,b)=\left[ \frac{Y_{l}\left( \mu b\right) }{g(k,\mu
,a,b)}-\frac{V(k,\mu ,\beta ,a,b)}{U(k,\mu ,\beta ,a,b)}\frac{J_{l}\left(
\mu b\right) }{g(k,\mu ,a,b)}\right] \times 
\end{equation*}%
\begin{equation*}
\frac{\left[ Y_{l}\left( \mu b\right) -\delta (k,\mu ,a,b)J_{l}\left( \mu
b\right) \right] }{J_{l}^{2}\left( kb\right) }
\end{equation*}%
Finally, the (GF) in this region $\left( r\leq b\right) $ is given by:%
\begin{equation}
G^{1,1}\left( l;r,r^{\prime }\right) =\left\{ 
\begin{array}{c}
\left[ Y_{l}\left( kr^{\prime }\right) -\left[ \psi (k,\mu ,\beta ,a,b)+%
\frac{Y_{l}\left( kb\right) }{J_{l}\left( kb\right) }\right] J_{l}\left(
kr^{\prime }\right) \right] J_{l}\left( kr\right) ,\ \ 0<r\leq r^{\prime }
\\ 
\left[ Y_{l}\left( kr\right) -\left[ \psi (k,\mu ,\beta ,a,b)+\frac{%
Y_{l}\left( kb\right) }{J_{l}\left( kb\right) }\right] J_{l}\left( kr\right) %
\right] J_{l}\left( kr^{\prime }\right) ,\ \ r^{\prime }\leq r\leq b%
\end{array}%
\right.   \label{a..20}
\end{equation}%
The resonance energies can be determined by the poles (see fig.2) of the
Green's function that is is to say by the poles of $\psi (k,\mu ,a,b)$ that
is to say 
\begin{equation*}
g(k,\mu ,a,b)=0
\end{equation*}%
or 
\begin{equation}
\gamma (k,\mu ,a,b)=\delta (k,\mu ,a,b)  \label{a223}
\end{equation}%
and from (\ref{116} - \ref{a111}) 
\begin{equation}
Y_{l}\left( \mu a\right) U(k,\mu ,\beta ,a,b)=J_{l}\left( \mu a\right)
V(k,\mu ,\beta ,a,b)  \label{a224}
\end{equation}%
and from (\ref{a1a} - \ref{a2a}) 
\begin{equation*}
\beta (k,\mu ,a,b)=\frac{Y_{l}(ka)}{J_{l}(ka)}
\end{equation*}%
By using (\ref{a33}) we find 
\begin{equation}
Y_{l}\left( ka\right) T(k,\mu ,a,b)=J_{l}\left( ka\right) F(k,\mu ,a,b)
\label{a225}
\end{equation}%
where $T(k,\mu ,a,b)$ and $F(k,\mu ,a,b)$ are defined above (\ref{a333} - %
\ref{a334}). Finally, Green's function in the region $\left( r>a\right) $ is
given by:%
\begin{equation*}
G^{3,3}\left( l;r,r^{\prime }\right) =-\left\{ 
\begin{array}{c}
J_{l}\left( kr^{\prime }\right) \left[ Y_{l}\left( kr\right) -\frac{F(k,\mu
,a,b)}{T(k,\mu ,a,b)}J_{l}\left( kr\right) \right] \quad \quad a\leq r\leq
r^{\prime } \\ 
\left[ Y_{l}\left( kr^{\prime }\right) -\frac{F(k,\mu ,a,b)}{T(k,\mu ,a,b)}%
J_{l}\left( kr^{\prime }\right) \right] J_{l}\left( kr\right) \quad \quad
r^{\prime }\leq r<\infty 
\end{array}%
\right. 
\end{equation*}

\section{Conclusion}

In this work, we have calculated the (GF) for the time-independent
Schrodinger equation in two dimensional space. The system considered in this
work is a quantum particle that have an energy $E$ and moves in an
axi-symmetrical potential. We have assumed that the Hamiltonian operator is
a piecewise continue operator: the potential $V(r)$, in which the quantum
particle moves, is equal to zero in the regions ($r<b$ and $r>a)$ and equal
a positive constant $V_{0}$ in a crown of internal radius $b$ and external
radius a ($b<a$). Our study focused on the diffusion states regime for which 
$E>V_{0}$ . We have used, to derive the (GF), the continuity of the solution
and of its first derivative at $r=b$ and $r=a$. We have obtained the
associate (GF) showing the resonance energies (for the case $E>V_{0}$).

\section{Acknowledgments}
 Each of us thanks the Editors to giving us the opportunity to publish our work
in this prestigious journal. Our thanks go also to the referees for their
valuable times that give for reading and commenting our work. We also thank
the dean of the LABTOP laboratory for supporting our work and substantial tools.

\bibliographystyle{siamplain}
\bibliography{references}

\address{
University of Hamma Lakhdar, Faculty of Exact Sciences, Eloued,39000, Algeria.\\
\email{benalibrahim@@ymail.com}}

\address{
University of Kasdi Merbah, FMSM faculty, LRPPS Laboratory, Ouargla,30000, Algeria.\\
\email{mewalid@@yahoo.com}}

\end{document}